\documentclass[fleqn,usenatbib]{mnras}

\usepackage{newtxtext,newtxmath}

\usepackage[T1]{fontenc}

\DeclareRobustCommand{\VAN}[3]{#2}
\let\VANthebibliography\thebibliography
\def\thebibliography{\DeclareRobustCommand{\VAN}[3]{##3}\VANthebibliography}

\usepackage{graphicx}	
\usepackage{amsmath}	

\title[Polar alignment of a circumbinary disc and planet]{Polar alignment of a circumbinary disc around a brown dwarf binary
}

\author[J. L. Smallwood et al.]{
Jeremy L. Smallwood,$^{1}$\thanks{E-mail: smallj2@ou.edu}
Thomas A. Baycroft$,^{2}$
Amaury H.M.J. Triaud$^{2}$
and
Richard P. Nelson$^{3}$
\\
$^{1}$Homer L. Dodge Department of Physics and Astronomy, The University of Oklahoma, Norman, OK 73019, USA\\
$^{2}$School of Physics and Astronomy, University of Birmingham, Edgbaston, Birmingham B15 2TT, UK\\
$^{3}$Astronomy Unit, Department of Physics and Astronomy, Queen Mary University of London, London E1 4NS, UK
}

\date{Accepted XXX. Received YYY; in original form ZZZ}

\pubyear{\the\year{}}

\begin{document}
\label{firstpage}
\pagerange{\pageref{firstpage}--\pageref{lastpage}}
\maketitle

\begin{abstract}
Inspired by recent observations suggesting that the retrograde precession of the brown dwarf binary 2M1510\,AB is consistent with induction by a polar circumbinary planet, we investigate the formation of such planets by studying the evolution of a primordial misaligned circumbinary disc around a brown dwarf binary. Analytical calculations show that a critical tilt angle of $i_{\rm crit} \gtrsim 50^\circ$ for moderately eccentric binaries is needed for polar alignment of circumbinary discs in systems with low disc-to-binary angular momentum ratios. For higher ratios, this angle converges to the Kozai-Lidov instability threshold of $\sim 39^\circ$. We identify disc parameters, such as viscosity ($\alpha = 10^{-4}$) and aspect ratio ($H/r = 0.05$), that enable polar alignment within typical disc lifetimes. Notably, a circumbinary disc around a low-mass binary, such as a brown dwarf binary, will require more time to achieve polar alignment compared to higher-mass systems. A hydrodynamical simulation confirms that an initially inclined disc around a brown dwarf evolves towards a polar state, creating favorable conditions for polar planet formation. Using these results, we finish by placing 2M1510\,AB into a wider context and speculate why such a polar circumbinary configuration has not been identified before.

\end{abstract}



\section{Introduction}
Binary brown dwarf systems are critical laboratories for studying the formation, evolution, and physical properties of substellar objects \citep{Stassun2006, Triaud2020}. Brown dwarfs occupy the mass range between stars and planets, typically between $\sim 13$ and 75 Jupiter masses, where core hydrogen fusion is insufficient to sustain long-term energy production \citep{Kumar1963,Chabrier2000,Burrows2001,Phillips2020}. The intermediate mass range makes them a unique bridge in our understanding of stellar and planetary formation processes. Furthermore, despite there being very few exoplanets known to orbit brown dwarfs, understanding their properties is a chance to investigate how the outcomes of planet formation scale with central mass.

Investigating brown dwarf multiplicity provides critical insights into their formation mechanisms by constraining key parameters such as orbital separation ($a_{\rm b}$) and mass ratio ($q = M_{\rm B}/M_{\rm A}$). For stellar binaries, orbital separations follow a log-normal distribution, with higher-mass stars exhibiting larger peak separations \cite[e.g.,][]{Raghavan2010,deRosa2014,Winterss2019}, while mass ratio distributions vary with primary mass, transitioning from bottom-heavy in O-stars to top-heavy in brown dwarfs \citep{Duchene2013,offner2022origin}. Brown dwarf binaries typically form tighter orbital configurations, with early studies of T0 to T8 spectral types \citep{Burgasser2003,Burgasser2006} identifying systems with separations under 5 au and mass ratios above 0.8. However, these studies were limited in sensitivity, primarily detecting systems with $q > 0.5$. Recent work has expanded the observed parameter space, revealing a broader range of mass ratios ($q = 0.2-1$)  and separations  \citep[$a = 0.1-1000\, \rm au$; e.g.][]{Fontanive2018}.

Recently, \citet{Triaud2020} reported the discovery of a fully substellar, hierarchical triple system in which the inner pair—2MASSW J1510478–281817 AB (hereafter 2M1510 AB), constitutes a double‐lined, eclipsing brown‐dwarf binary, while the outer tertiary, 2MASS J15104761–2818234 (hereafter 2M1510 C), resides at a projected separation of $\sim$250 AU. Both components were originally catalogued in the Two Micron All‐Sky Survey \citep{Cutri2003}. High‐precision photometric and spectroscopic monitoring of 2M1510 AB yields an orbital period $P_{\rm orb}\approx20$ days and an orbital eccentricity $e=0.36\pm0.02$ \citep{Triaud2020}. At a mean Gaia DR2 parallax distance of $36.6\pm0.3\, \rm pc$ \citep{Bailer-Jones2018}, the tertiary’s projected separation corresponds to $\sim250\, \rm au$. Kinematic analysis—combining Gaia proper motions, parallax, and ground‐based radial velocities—confirms that all three substellar components are co‐moving members of the $\sim45\pm5,$Myr Argus moving group \citep{Gagne2015,Zuckerman2019,Triaud2020}.  Complementary observations by \citet{Calissendorff2019} uncovered a fourth, non-eclipsing companion (designated 2M1510 D) with an inferred mass of $M_{\rm D}\approx17.7^{+4.2}_{-2.1}\, \rm M{\rm Jup}$ on a $\sim4.4\, \rm au$ orbit about the AB pair. However, 2M1510 D was observed at only a single epoch. Constraining its proper motion requires observations at two or more epochs, and future observations are needed to confirm whether this companion candidate is gravitationally bound to the 2M1510 system.

\cite{Baycroft2025} performed radial velocity measurements of the brown dwarf components, revealing that the binary undergoes retrograde apsidal precession. They propose this precession is induced by the gravitational influence of a circumbinary planet in a near polar orbital configuration. For such a polar planet to exist around 2M1510 AB, a primordial disc would need to have formed initially misaligned with the binary orbital plane and subsequently evolved into a near-polar configuration before planet formation commenced \citep[as in][]{Smallwood2020a}. Polar-aligned circumbinary discs are thought to provide a favorable environment for the formation of polar planets \citep{Smallwood2024b, Smallwood2024c}. Despite the lack of confirmed polar circumbinary planets, there is indirect evidence of such an orbital configuration in the post–asymptotic giant branch star binary AC Her \citep{Martin2023}.

Besides the claimed highly inclined planet 2M1510 (AB)b, there is also one example of a potentially misaligned planetary-mass object orbiting a brown dwarf binary. VHS 1256 (AB)b is a directly imaged planetary-mass companion orbiting a binary brown dwarf, VHS 1256 AB \citep[total mass $\sim 0.141 \pm 0.008\, \rm M_{\odot}$][]{Dupuy2023}, in a hierarchical triple system \citep{Gauza2015,Rich2016,Stone2016}. VHS 1256 (AB)b has an estimated mass of $12.0 \pm 0.1\, \rm M_{\rm jup}$ or $16.0 \pm 1\, \rm M_{\rm jup}$ \citep{Dupuy2023,Miles2023}, with a separation of $\sim 350^{+110}_{-150}\, \rm au$ and an estimated orbital eccentricity of $0.7 \pm 0.1$. The separation and eccentricity of the binary are $1.96 \pm 0.03\, \rm au$ and $0.883 \pm 0.003$, respectively, with a system age of $140 \pm 20\, \rm Myr$. \citet{Poon2024} estimated the inclination of VHS 1256 (AB)b with respect to the binary orbital plane to be $\sim 115\pm14^{\circ}$ \citep{Dupuy2023}, which is close to a polar orbit. 

While no circumbinary discs around brown dwarf binaries have been directly observed, simulations suggest they can form and accrete material under certain conditions \cite[e.g.,][]{Dittmann2024,Bate2003}. Circumbinary discs are observed at various stages of stellar evolution \cite[e.g.,][]{Czekala2019}, implying that such discs around brown dwarf binaries are plausible and likely share properties with those around single brown dwarfs. Brown dwarf discs, initially detected in near- and mid-infrared surveys \citep{Comeron1998,Muench2001,Natta2001,Natta2002,Jayawardhana2003}, are now studied across infrared, submillimeter, and millimeter wavelengths \citep{Testi2016,Daemgen2016}. These discs resemble those around classical T Tauri stars, exhibiting similar diversity in geometry and dust properties \citep{Mohanty2004}. Evidence of millimeter-sized dust grains in brown dwarf discs has been identified \cite[e.g.,][]{Apai2005,Scholz2007}, with high-resolution ALMA observations confirming substantial dust grain growth \citep{Ricci2012,Ricci2014}. 

In this letter, we investigate for the first time how the polar alignment of a circumbinary disc around a brown dwarf binary differs from that around a solar-type binary. We apply our findings to the system 2M1510\,AB, which shows potential signatures of a polar circumbinary planet. If such a planet exists, it would likely have formed from a primordial polar-aligned circumbinary disc around the system. We constrain the initial disc properties required for polar alignment to occur. The outline of the letter is as follows: In Section~\ref{sec::timescales}, we discuss the criteria and timescales for polar alignment around a brown dwarf binary. In Section~\ref{sec::methods}, we detail the methods used for the hydrodynamical simulations. In Section~\ref{sec::hydro_results}, we present the results of the hydrodynamical simulations of a tilted circumbinary disc around a brown dwarf binary. In Section~\ref{sec::discussion}, we discuss the presence of the outer companion 2M1510\,C , the efficiency of planet formation around brown dwarfs, and place the 2M1510\,AB system into the wider context of circumbinary exoplanet detections. Finally, we summarise our conclusions in Section~\ref{sec::conclusion}.

\section{Circumbinary disc alignment timescales}
\label{sec::timescales}
Disc misalignment can occur at various stages of stellar evolution and arise from several processes. In star-forming regions, misaligned discs may result from the chaotic accretion of turbulent molecular clouds, introducing random angular momentum directions \citep{Offner2010, Bate2012, Tokuda2014}. Post-formation accretion of misaligned gas can further alter the orientation of a circumbinary disc relative to the binary plane \citep{Bate2010, Bate2018}. Additionally, if a binary forms in an elongated cloud, its orbital plane can be misaligned with the cloud's rotation axis due to anisotropic collapse or dynamical interactions \citep{Bonnell1992}. A misaligned circumbinary disc will undergo alignment process evolving towards a coplanar or polar configuration \cite[e.g.,][]{Papaloizou1995,Lubow2000,Aly2015,Martin2017,Zanazzi2018}. 

Within the linear warp propagation theory, the warp is dissipated and a initially tilted circumbinary disc polar aligns on a timescale
\begin{equation}
    \tau_{\rm polar} = \frac{1}{\alpha}\bigg( \frac{H}{r}\bigg)^2 \frac{\Omega_{\rm b}}{\Omega_{\rm d}^2},
    \label{eq::tau}
\end{equation}
where $\alpha$ is the disc viscosity, $H/r$ is the disc aspect ratio, and $\Omega_{\rm b} = \sqrt{G(M_1 + M_2)/a^3}$ is the binary angular frequency. $\Omega_{\rm d}$ is the global disc precession frequency, which differs depending on coplanar or polar alignment, and is given by
\begin{equation}
  \Omega_{\rm d}=
    \frac{3\sqrt{5}}{4}e_{\rm b}\sqrt{1+4e_{\rm b}^2}\frac{M_1M_2}{M^2}\left<\left( \frac{a}{r}\right)^{7/2}\right>\Omega_{\rm b},
  \label{eq::omega}
\end{equation}
\citep{Lubow2018,Smallwood2019},
where
\begin{equation}
    \left<\left( \frac{a}{r}\right)^{7/2}\right>=\frac{\int_{r_{\rm in}}^{r_{\rm out}}\Sigma r^3 \Omega (a/r)^{7/2}\,dr }{\int_{r_{\rm in}}^{r_{\rm out}} \Sigma r^3 \Omega \, dr}.
    \label{eq::ar}
\end{equation}
Note that if the disc breaks, the constituent discs will precess at different rates and Eq.~\ref{eq::ar} cannot be applied. 

\begin{figure*} \centering
\includegraphics[width=0.68\columnwidth]{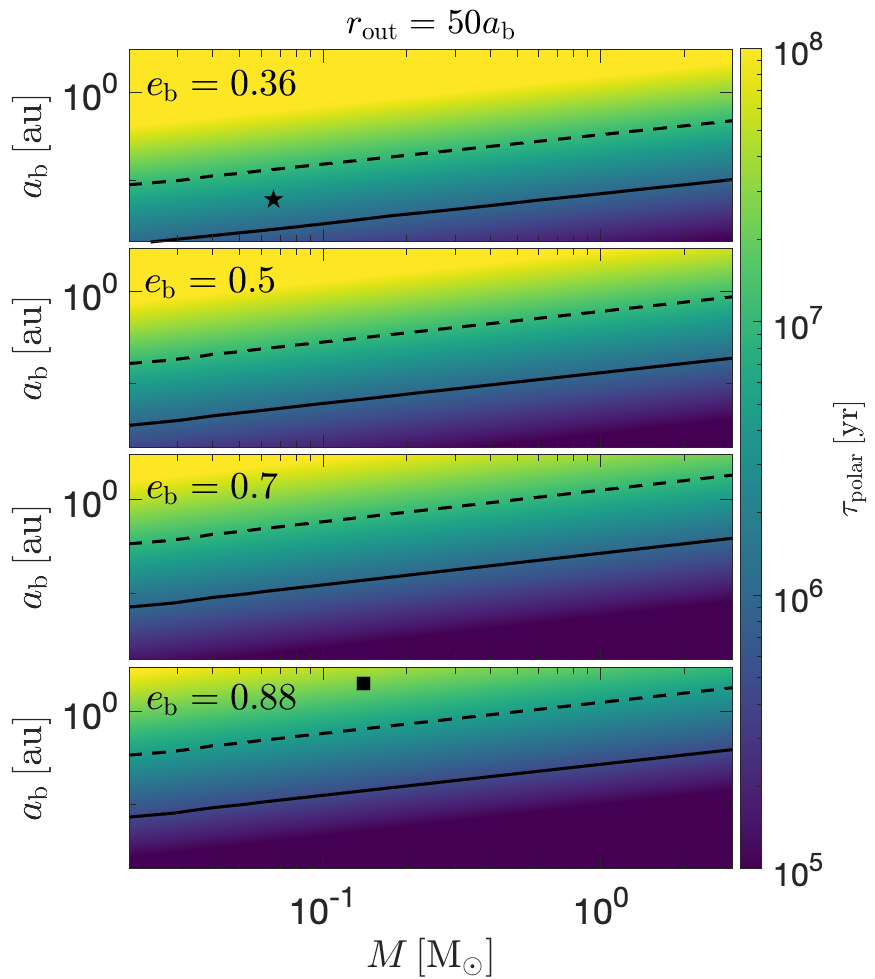}
\includegraphics[width=0.68\columnwidth]{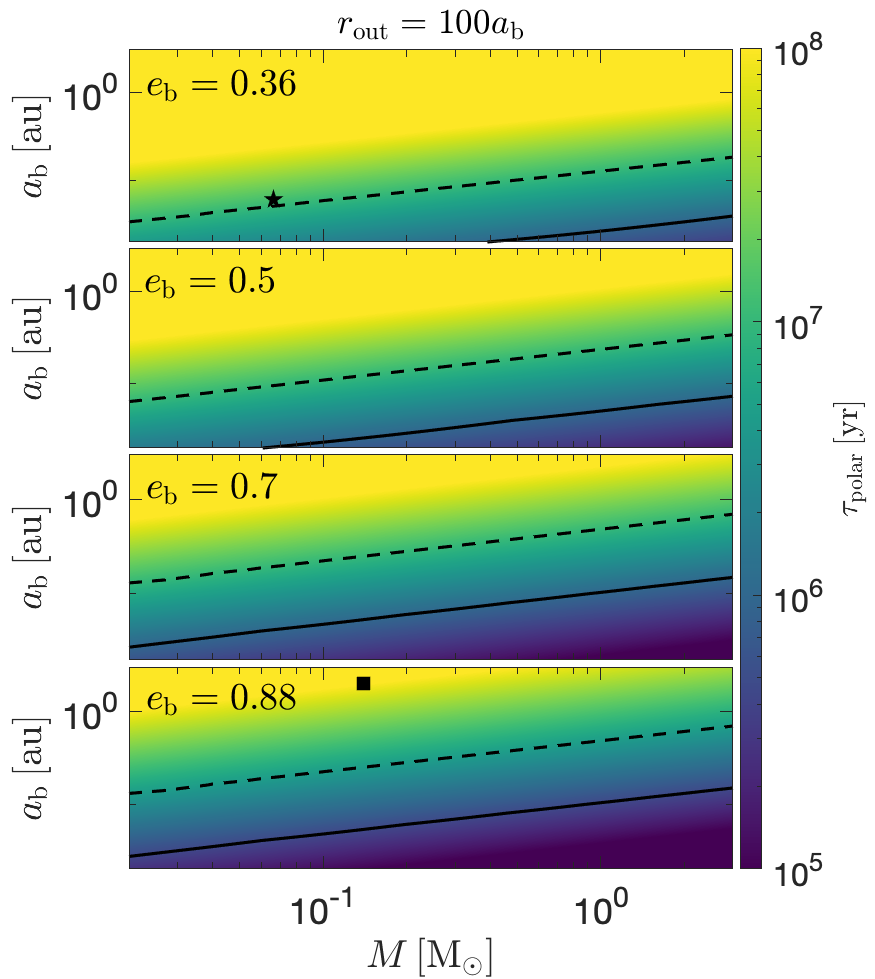}
\includegraphics[width=0.68\columnwidth]{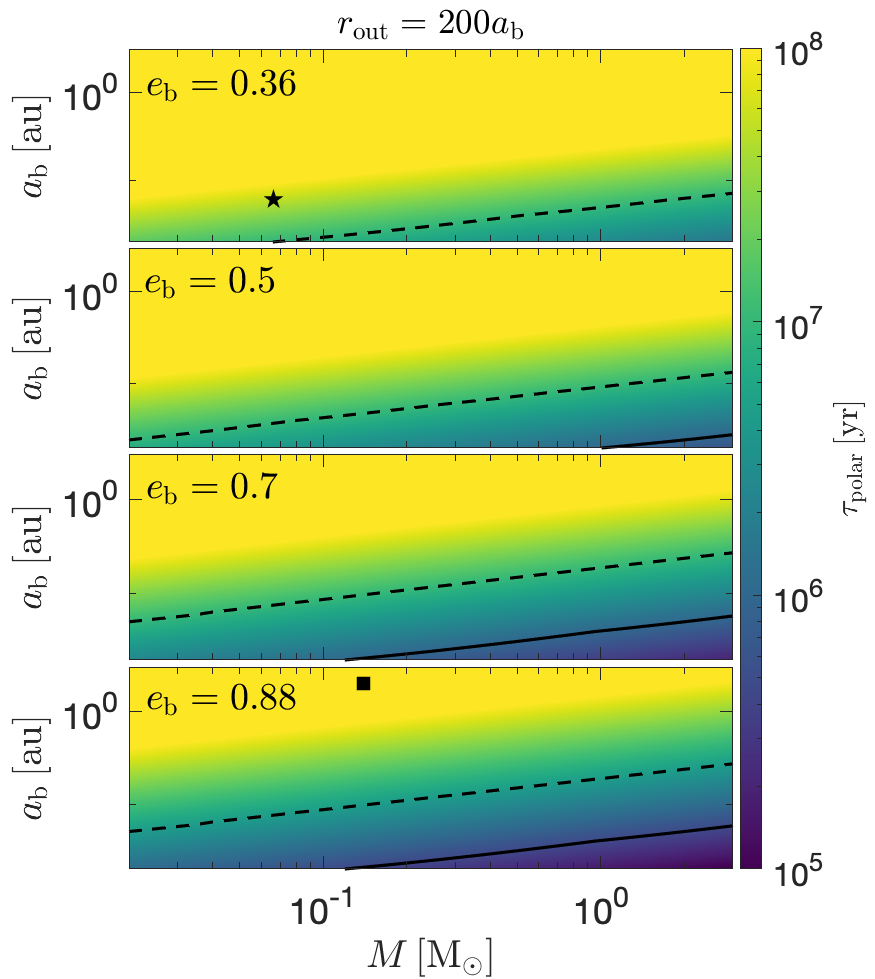}
\caption{Polar alignment timescale ($\tau_{\rm polar}$) from Eq.~(\ref{eq::tau}) as a function of binary mass ($M$) and semi-major axis ($a_{\rm b}$) for binary eccentricities $e_{\rm b} = 0.36$,  $e_{\rm b} = 0.5$, $e_{\rm b} = 0.7$, and $e_{\rm b} = 0.88$ (from top to bottom), for an outer disc radius $r_{\rm out} = 50a_{\rm b}$ (left plot), $r_{\rm out} = 100a_{\rm b}$ (middle plot), and  $r_{\rm out} = 200a_{\rm b}$ (right plot).  Higher masses and smaller separations lead to faster alignment, with shorter timescales at higher eccentricities.  We overlay the estimated lifetime of a protoplanetary disc, ranging from $10^6\, \rm yr$ (solid line) to $10^7\, \rm yr$ (dashed line). The star and square marks the approximate parameters for 2M1510\,AB and VHS 1256 AB, respectively. }
\label{fig::alignment_time} 
\end{figure*}

Figure~\ref{fig::alignment_time} demonstrates how the polar alignment timescale ($\tau_{\rm polar}$) from Eq.~(\ref{eq::tau}) varies with total binary mass ($M$), semi-major axis ($a_{\rm b}$), and eccentricity ($e_{\rm b}$). For these calculations, we assume an equal-mass binary ($M_1 = M_2$), $\alpha = 10^{-4}$, and $H/r = 0.05$, $r_{\rm in} = 2a_{\rm b}$. We show the results for $r_{\rm out} = 50a_{\rm b}$ (left plot), $r_{\rm out} = 100a_{\rm b}$ (middle plot), and $r_{\rm out} = 200a_{\rm b}$ (right plot). Across all panels, the alignment timescale decreases with increasing binary mass, indicating that discs around more massive systems align more quickly. Similarly, alignment times increase with larger semi-major axis, suggesting that discs around wide binary systems take longer to align, which is consistent with observations \cite[e.g.,][]{Czekala2019}. Also, a wider disc will take longer to align than a more truncated disc. The four panels compare binary eccentricities $e_{\rm b} = 0.36$, $e_{\rm b} = 0.5$, $e_{\rm b} = 0.7$, and $e_{\rm b} = 0.88$, revealing that higher binary eccentricities correspond to shorter disc alignment times across the parameter space. This indicates that more eccentric binary systems promote faster alignment, while lower eccentricities result in slower polar alignment. 

In Figure~\ref{fig::alignment_time}, we also overlay the estimated lifetime of a protoplanetary disc, ranging from $10^6\, \rm yr$ (solid line) to $10^7\, \rm yr$ (dashed line). The star marks the parameters for 2M1510\,AB, indicating that a circumbinary disc around a low-mass binary, such as a brown dwarf binary, will require more time to achieve polar alignment compared to higher-mass systems. Nevertheless, for typical protoplanetary disc parameters, the disc is expected to align within a typical 10 Myr lifetime given that the radial disc extent isn't too large. 

The square represents the parameters for VHS 1256\,AB (which orbits at approximately $200a_{\rm b}$), indicating that the alignment timescale exceeds the disc lifetime for both $r_{\rm out} = 50a_{\rm b}$, $r_{\rm out} = 100a_{\rm b}$, and $r_{\rm out} = 200a_{\rm b}$, despite the high binary eccentricity ($e_{\rm b} = 0.88$). This implies the nearly polar planetary-mass companion, VHS 1256 b, is qualitatively different from 2M1510(AB)b, and either formed from a primordially near-polar disc, or more likely that it acquired its highly inclined orbit  through a scattering event, which is supported by its significant eccentricity ($e \approx 0.7$) whereas a polar circumbinary disc is expected to have less eccentricity growth \citep{Smallwood2022}.

\begin{figure} \centering
\includegraphics[width=1\columnwidth]{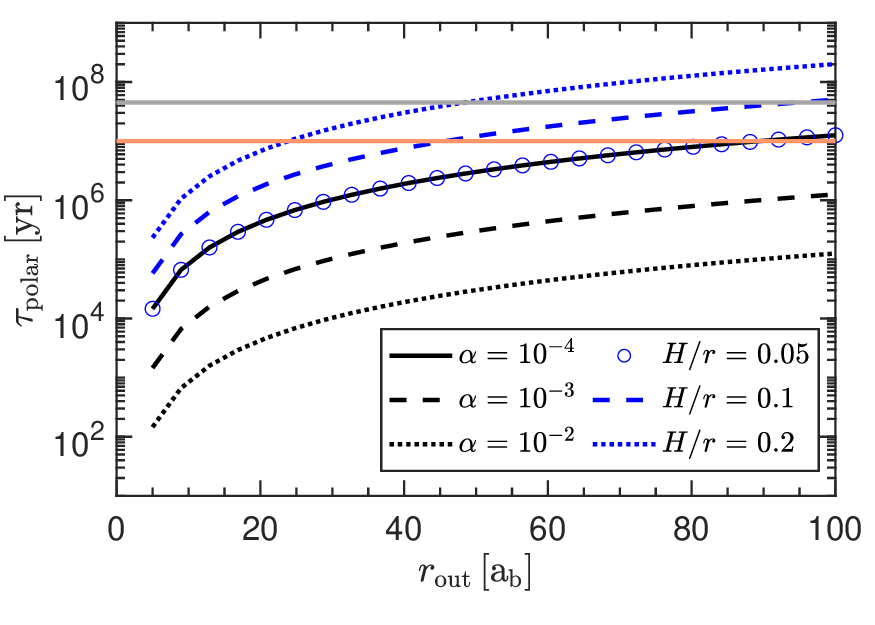}
\caption{The polar alignment timescale, $\tau_{\rm polar}$, is shown as a function of the circumbinary disc radius, $r_{\rm out}$, expressed in units of the binary semi-major axis, $a_{\rm b}$, for different values of disc viscosity (black) with disc aspect ratio $H/r = 0.05$, and different values of $H/r$ (blue) with $\alpha = 10^{-4}$. The horizontal red line represents the estimated upper limit for the age of a protoplanetary disc, which ranges from $1$ to $10\, \rm Myr$. The horizontal gray line indicates the estimated age of 2M1510\,AB ($\sim 45\, \rm Myr$). }
\label{fig::align}
\end{figure}

We now examine the conditions necessary for a polar-aligned circumbinary disc around 2M1510\,AB. Figure~\ref{fig::align} illustrates the polar alignment timescale, $\tau_{\rm polar}$, as a function of the circumbinary disc outer radius, $r_{\rm out}$, expressed in units of the binary semi-major axis, $a_{\rm b}$. The figure presents a family of curves representing different values of disc viscosity ($\alpha$) and disc aspect ratio ($H/r$). The black curves show the effect of varying $\alpha$ for a fixed $H/r = 0.05$, while the blue curves demonstrate the influence of changing $H/r$ for a constant $\alpha = 10^{-4}$. For a fixed disc aspect ratio, increasing the disc viscosity results in a faster polar alignment timescale. This implies that more viscous discs are more efficient at aligning perpendicular with the binary. For a fixed viscosity, increasing the disc aspect ratio leads to a longer polar alignment timescale. The horizontal red line represents the estimated upper limit for the age of a protoplanetary disc, which ranges from $1$ to $10\, \rm Myr$. Brown dwarf inner disc lifetimes do not appear to be vastly different from those of discs around T Tauri stars \cite[e.g.,][]{Jayawardhana2003}. This line provides a reference point to evaluate whether the disc in this system has had enough time to reach a polar state before the disc disperses. However, circumbinary discs may have longer lifetimes than discs around single stars. For example, the discs around V4046 Sgr and AK Sco have ages of $23 \pm 3, \rm Myr$ \citep{Mamajek2014} and $18 \pm 1, \rm Myr$ \citep{Czekala2015}, respectively. The horizontal gray line represents the estimated age of 2M1510\,AB ($\sim 45\, \rm Myr$).

\begin{figure} \centering
\includegraphics[width=1\columnwidth]{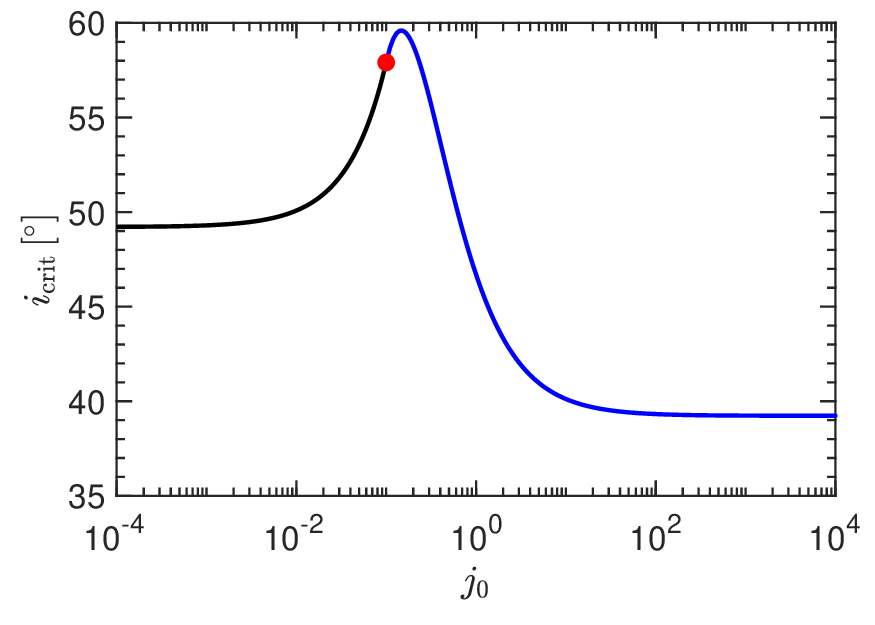}
\caption{The critical tilt separating coplanar versus polar alignment from Eq.~(\ref{eq::icrit}) as a function of disc angular momentum to binary angular momentum ratio, $j_0$. The black and blue curves represent to low $j_0$ and high $j_0$ approximations, respectively. The red dot is the transition point between these two regimes.}
\label{fig::icrit}
\end{figure}

 The critical tilt, $i_{\rm crit}$, for a circumbinary disc's angular momentum vector to align with the binary angular momentum (coplanar alignment) or to the binary eccentricity vector (polar alignment) depends on the binary eccentricity, $e_{\rm b}$, and the ratio between the initial disc angular momentum to the initial binary angular momentum, $j_0 = J_{\rm d0}/J_{\rm b0}$. The analytical approximate for $i_{\rm crit}$ is estimated given two regimes shown by
\begin{equation}
   i_{\rm crit}=\begin{cases}
    \arccos \bigg[ \frac{\sqrt{5}e_{\rm b} \sqrt{4e_{\rm b}^2 -4j_0^2 (1-e_{\rm b}^2)+1}-2j_0(1-e_{\rm b}^2)}{1+4e_{\rm b}^2} \bigg], & \text{low $j_0$},\\
     \arccos \bigg[ \frac{\sqrt{(1-e_{\rm b}^2)(1+4e_{\rm b}^2) + 60(1-e_{\rm b}^2)j_0^2}-(1-e_{\rm b}^2)}{10(1-e_{\rm b}^2)j_0} \bigg], & \text{high $j_0$},
  \end{cases}
  \label{eq::icrit}
\end{equation}
\citep{Martin2019}. The initial angular momentum of the binary is given by
\begin{equation}
    J_{\rm b0} = \mu\sqrt{G(M_1 + M_2)a_{\rm b0}(1-e_{\rm b0}^2)},
\end{equation}
where $\mu $ is the reduced mass and the initial angular momentum of the disc is given by
\begin{equation}
    J_{\rm d0} = \int_{r_{\rm in}}^{r_{\rm out}} 2\pi r^3\Sigma_0(r)\Omega dr,
\end{equation}
where $\Omega = \sqrt{GM/r^3}$ is the Keplerian angular velocity. We explore a range of $j_0$ from $10^{-4}$ to $10^{4}$ using the observed binary eccentricity of 2M1510\,AB, $e_{\rm b} = 0.36$. Figure~\ref{fig::icrit} shows the critical tilt angle, $i_{\rm crit}$, from Eq.~(\ref{eq::icrit}) as a function of $j_0$. For low values of $j_0$, the critical tilt is approximately $50^\circ$, while for high $j_0$, $i_{\rm crit}$ approaches $40^\circ$. In the limit of large $j_0$, the critical tilt approaches the critical angle for Kozai–Lidov oscillations \citep{Martin2019}. The transition between these regimes occurs at $j_0 = 0.1$. The peak in $i_{\rm crit}$ occurs at $j_0\sim 0.14$ with $i_{\rm crit} \sim 60^\circ$.

\section{Hydrodynamical simulations}
\label{sec::methods}

\subsection{Setup}
As a proof-of-concept, we use the smoothed particle hydrodynamics (SPH) code {\sc phantom} \citep{Price2018} to simulate an initially inclined circumbinary disc around the brown dwarf binary 2M1510\,AB. The discs within our hydrodynamical model are in the bending wave regime, meaning that the disc aspect ratio, $H/r$, is larger than the \cite{Shakura1973} $\alpha$-viscosity parameter, which is appropriate for protoplanetary discs \cite[e.g.,][]{Hueso2005,Rafikov2016,Ansdell2018}. In this regime, warps induced in the disc by the brown dwarf binary torque propagate as  bending waves with a vertical averaged speed $c_{\rm s}/2$ \citep{Papaloizou1995}, where $c_{\rm s}$ is the sound speed.

We use the observed binary parameters of 2M1510\,AB from \cite{Triaud2020} and \cite{Baycroft2025}, with $a_{\rm b} = 0.06\, \rm au$ (semi-major axis), $e_{\rm b} = 0.36$ (eccentricity), $M_1 = 0.033104\, \rm M_\odot$ (primary mass) and $M_2 = 0.033219\, \rm M_\odot$ (secondary mass) . The binary separation and eccentricity are allowed to evolve in time. We employ a Cartesian coordinate system ($x$,$y$,$z$), where the $x$-axis aligns with the direction of the binary eccentricity vector, and the $z$-axis aligns with the direction of the binary angular momentum vector. The binary accretion radii are set to $r_{\rm acc,1} = r_{\rm acc,2} = 0.03\, \rm au$, in order to increase computational efficiency. The sink accretion radius is considered a hard boundary, where the accreted particles' mass and angular momentum are added to the sink \citep{Bate1995}. 

Observations of brown dwarfs in the (sub)-mm and millimeter wavelengths indicate that disc masses are roughly proportional to stellar mass, typically around $1\%$ of the stellar mass, although with considerable scatter \citep{Scholz2006,Klein2003,Mohanty2013,Andrews2013}. In our models, the circumbinary disc is represented by $1 \times 10^6$ equal-mass SPH particles, with a total disc mass of $0.1\%$ of the total binary mass. This simplification ensures that the disc mass does not significantly affect the orbit of the binary. A more massive circumbinary disc will align to a stationary tilt with respect to the binary \citep{Martin2019}. The particles are radially distributed from the inner disc radius, $r_{\rm out} = 0.12\, \rm au\, (\sim 2a_{\rm b})$, to the outer disc radius, $r_{\rm out} = 0.6\, \rm au\, (\sim 10a_{\rm b})$. We adopt a narrow, rigidly precessing disc configuration to avoid dynamical instabilities such as disc breaking or tearing, which arise when the radial communication timescale is longer than the precession timescale \citep{Nixon2012, Facchini2013}. The circumbinary disc is initially misaligned by $75^\circ$, causing it to evolve toward a polar state (see Fig.~\ref{fig::icrit}).

The gas surface density profile is initially a power-law distribution given by $ \Sigma(r) = \Sigma_0 (r/r_{\rm in})^{-p}$, where $\Sigma_0 = 1.62\times10^{3}\, \rm g/cm^2$ is the density normalization (defined by the total mass), $p$ is the power law index, and $r$ is the spherical radius. We set $p=3/2$. We adopt the locally isothermal equation of state. The disc thickness is scaled with radius as $ H = c_{\rm s}/\Omega \propto r^{3/2-q}$, where $\Omega = \sqrt{GM/r^3}$ and $q = 3/4$.  We set an initial gas disc aspect ratio of $H/r = 0.05$ at $r = r_{\rm in}$. The \cite{Shakura1973} viscosity, $\alpha_{\rm SS}$, prescription is given by $\nu = \alpha_{\rm SS} c_{\rm s} H$, where $\nu$ is the kinematic viscosity. In order to simulate
$\alpha_{\rm SS}$, we use the artificial viscosity $\alpha^{\rm av}$ prescription in \cite{Lodato2010} given as $\alpha_{\rm SS} \approx (\alpha_{\rm AV}/10)(\langle h \rangle/H)$.
We take the \cite{Shakura1973} $\alpha_{\rm SS}$ parameter to be $0.005$. The circumbinary disc is initially resolved with average smoothing length per scale height $\langle h \rangle / H = 0.34$. With the above prescription,  $\langle h \rangle / H$ and $\alpha_{\rm SS}$ are constant over the radial extent of the disc \citep{Lodato2007}.  Note, that $\langle h \rangle / H$ can vary due to the formation of warps in the disc \citep{Fairbairn2021,Deng2022}.

We separate the disc into 300 bins in spherical radius. Within each bin, we calculate the azimuthally averaged  tilt and longitude of ascending node.

\begin{figure} \centering
\includegraphics[width=0.49\columnwidth]{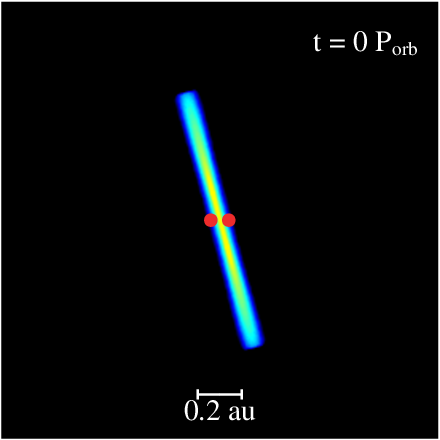}
\includegraphics[width=0.49\columnwidth]{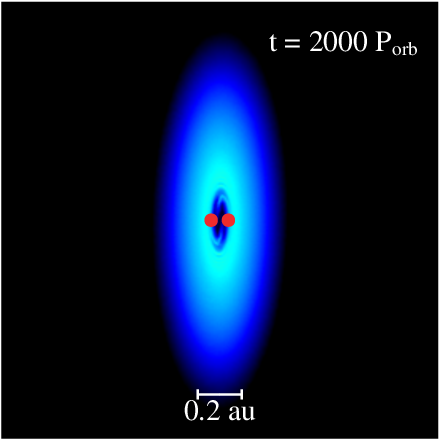}
\includegraphics[width=\columnwidth]{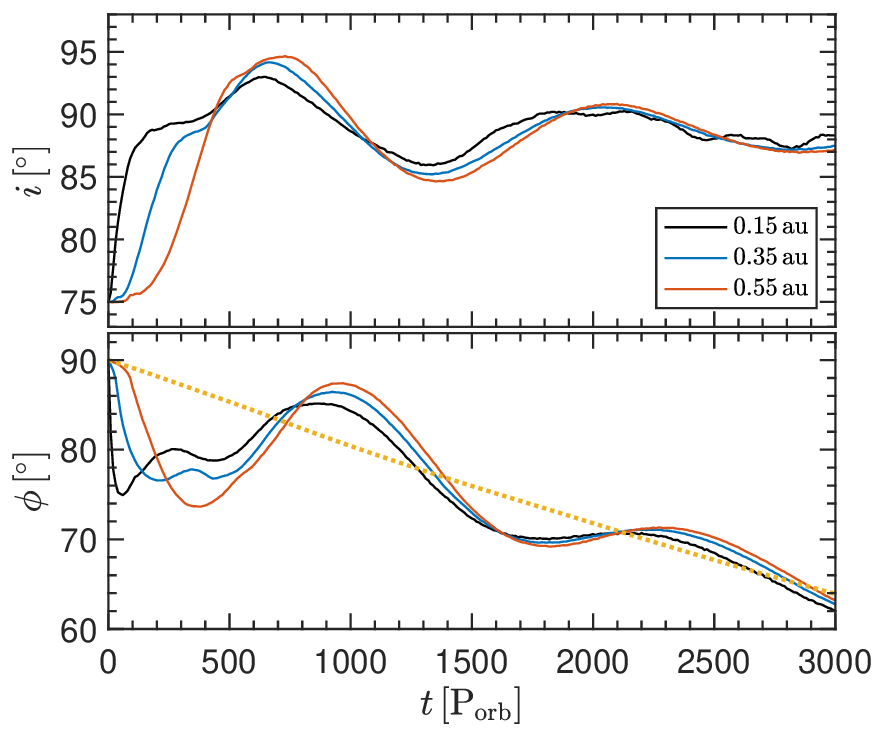}
\caption{Top left panel: Initial disc structure tilted by $75^\circ$, with the color indicating surface density—yellow corresponds to values approximately two orders of magnitude greater than blue. The brown dwarf binary components are represented by red dots, and the disc is viewed in the $x$–$z$ plane. Top right panel: Disc structure at $t = 1500, \rm P_{orb}$. Bottom panel: Evolution of the disc tilt, $i$, and longitude of the ascending node, $\phi$, over time. The disc is probed at three different radii: $0.15, \rm au$ (black), $0.35, \rm au$ (blue), and $0.55, \rm au$ (red), corresponding to the inner, middle, and outer regions of the disc. The dotted yellow line represents the azimuthal angle of the binary's eccentricity vector. The circumbinary disc is undergoing alignment to a polar configuration relative to the brown dwarf binary. }
\label{fig::hydro}
\end{figure}

\subsection{Results}
\label{sec::hydro_results}

Figure~\ref{fig::hydro} illustrates the evolution of a circumbinary disc tilted initially by $75^\circ$ relative to the orbital plane of a brown dwarf binary. The top left panel depicts the initial structure of the disc, where the color scale represents the surface density, with yellow regions indicating values approximately two orders of magnitude higher than the blue regions. The binary components are shown as red dots, and the system is viewed in the $x$--$z$ plane. The top right panel shows the disc's structure at $t = 1500, \rm P_{orb}$, where the disc continues the precess about the eccentricity vector of the binary and is transitioning toward polar alignment.

The bottom panels display the time evolution of the disc's tilt, $i$, and the longitude of the ascending node, $\phi$, at three distinct radii: $0.15, \rm au$ (black), $0.35, \rm au$ (blue), and $0.55, \rm au$ (red). These radii correspond to the inner, middle, and outer regions of the disc, respectively. The disc tilt, $i$, begins at $75^\circ$ and evolves over time, with tilt oscillations driven by the eccentricity of the binary \cite[e.g.,][]{Smallwood2019}. Since we model a moderately extended disc, the inner disc region (black curve) aligns more rapidly compared to the middle (blue) and outer (red) regions, highlighting differential alignment across the disc.

The evolution of $\phi$ indicates the disc precession. The dotted yellow line in the bottom panel represents the azimuthal angle of the binary's eccentricity vector. The disc's longitude of the ascending node, $\phi$, converges toward the binary's azimuthal angle, further confirming the polar alignment process. Over time, the entire circumbinary disc transitions toward a polar configuration relative to the binary's orbital plane.

The hydrodynamical simulation demonstrates that despite its overall low mass, a binary brown dwarf system, such as 2M1510\,AB, remains highly efficient in polar-aligning a circumbinary disc. The results indicate that the timescale for polar alignment is shorter than the timescale required for substantial grain growth within the disc \cite[e.g.,][]{Smallwood2024b,Smallwood2024c}. This implies that the dust will remain well-coupled to the gas as the disc aligns to a polar orientation relative to the binary's orbital plane. The efficient polar disc alignment has important implications for planet formation. A polar-aligned disc provides the conditions necessary for the formation of polar circumbinary planets, as the dust and gas dynamics facilitate the accumulation and growth of planetesimals in a stable polar configuration \citep{Smallwood2024c}. Such a process could explain the potential existence of a polar-aligned planet around 2M1510\,AB, offering a unique environment for planetary system formation compared to the more common coplanar circumbinary systems.

\section{Discussion}
\label{sec::discussion}

\subsection{The presence of 2M1510 C}
 Based on \cite{Martin2022}, a circumbinary disc can still evolve into a polar alignment despite the presence of an outer companion. The companion may excite von Zeipel-Kozai-Lidov oscillations \citep{vonZeipel:1910nh,Kozai1962,Lidov1962}, which could reduce the disc misalignment if the companion's torque dominates. Conversely, if the inner binary's torque dominates, the disc will align to a polar orientation. An example is HD 98800 BaBb, which hosts a polar-aligned circumbinary disc \citep{Kennedy2019}. This system belongs to a hierarchical quadruple star system, where the A-B semi-major axis is only $54\, \rm au$ ($\sim 54$ times the separation of the B binary). The circumbinary disc maintains its polar alignment despite the proximity of the outer binary companion and remains stable over long timescales without being susceptible to Kozai-Lidov (KL) oscillations \cite[e.g.,][]{Smallwood2022,Martin2022}. Given that the separation of 2M1510 C is $\sim 250\, \rm au$ ($\sim4200$ times the separation of the AB binary), the impact of 2M1510 C  on the polar alignment of the disc around 2M1510 AB will therefore be negligible.

The vZLK oscillation of a circumbinary disc can be quenched by gas‐pressure–induced precession when the dimensionless stability parameter $S$ falls below unity \citep{Zanazzi2017}:
\begin{equation}
    S = 0.36 \bigg( \frac{a_{\rm AB-C}}{3R_{\rm out}}\bigg)^3 
    \frac{M_{\rm AB}}{M_{\rm C}} \bigg(\frac{H(R_{\rm out})}{0.1R_{\rm out}}\bigg)^2 \lesssim 1, 
    \label{eq::S_value}
\end{equation}
where $a_{\rm AB-C}$ is the separation of the AB-C binary, $M_{\rm AB}/M_{\rm C}$ is the mass ratio of the central binary with the outer companion, $H (R_{\rm out})$ is the disc scale-height at the outer disc edge. Physically, $S^{-1}$ measures the ratio of the external companion’s tidal torque on the disc to the internal torque arising from gas pressure; when gas‐pressure torques dominate ($S \ll 1$), vZKL oscillations are suppressed \citep{Zanazzi2017,Nealon2025}. For a given choice of $R_{\rm out}$, we compute $H$ using the  irradiated, flared‐disc prescription of \citet{Chiang1997}, adopting the known luminosity of the central AB binary \citep{Triaud2020}. 

\begin{figure} \centering
\includegraphics[width=1\columnwidth]{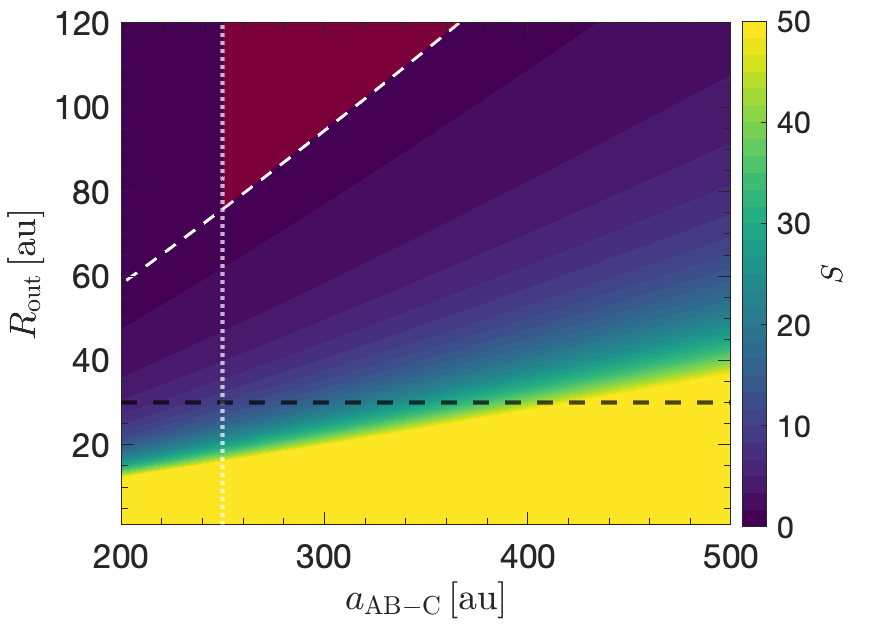}
\caption{The $S$-parameter from Eq.~(\ref{eq::S_value}) is shown as a function of the binary separation, $a_{\rm AB-C}$, and the outer edge of the gaseous disc, $R_{\rm out}$. The dashed curve represents the points where $S \lesssim 1$. The dotted line corresponds to the lower limit of the projected separation of 2M1510 AB, $250\, {\rm au}$. The red-shaded region indicates the combinations of $a_{\rm b}$ and $R_{\rm out}$ for which the disc is susceptible to the vZLK instability.}
\label{fig::ZKL_pcolor}
\end{figure}

Figure~\ref{fig::ZKL_pcolor} maps the stability parameter $S$ for triggering the vZLK instability as a function of the binary semi-major axis $a_{\rm AB-C}$ and the gaseous disc's outer radius $R_{\rm out}$. The red‐shaded region delineates the ($a_{\rm AB-C}$,$R_{\rm out}$) combinations that permit vZLK oscillations. Given the low mass of the outer companion, the radial extent of the disc must be sufficiently large compared to the AB binary separation for the instability to occur. For an assumed AB–C separation of $a_{\rm AB-C} = 250\, \rm au$, we infer a minimum outer edge of $R_{\rm out} \approx 75\, \rm au$. We emphasize that the observed projected separation of $250\, \rm au$ likely underestimates the true semi‑major axis; any increase in correspondingly raise the required $R_{\rm out}$, since vZKL torques must overcome gas-pressure-induced precession to remain operative.

\begin{figure} \centering
\includegraphics[width=1\columnwidth]{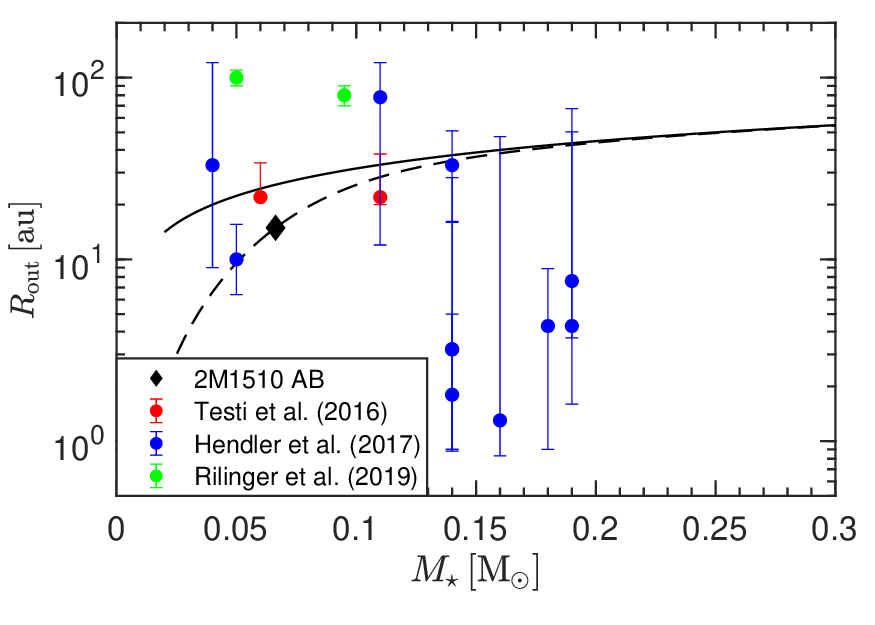}
\caption{ The observed outer disc edge, $R_{\rm out}$, as a function of star mass of discs around brown dwarfs from \citet{Testi2016} (red), \citet{Hendler2017} (blue), and \citet{Rilinger2019} (green). The solid black curves denotes the scaling relation from Eq.~\ref{eq::r_out} and the dashed black curve denotes the modified scaling relation from Eq.~\ref{eq::r_out_mod}. The black diamond represent the estimated radial extent of a primordial disc around 2M1510 AB.}
\label{fig::disc_size}
\end{figure}

Given that our analysis depends critically on the radial extent of the disc around 2M1510 AB, we estimate the size of its primordial disc from the total mass of the central binary. Observations indicate that discs around very low‐mass stars and brown dwarfs flatten relative to a simple power‐law scaling at low masses: \cite{Hendler2017} show that these discs are smaller than expected from a linear size–mass relation, and \cite{Mohanty2013} find typical outer radii of only $\sim 20\, \rm au$ for brown‐dwarf discs. The canonical empirical relation between stellar mass $M_\star$ and dust-disc outer radius $R_{\rm out}$ is 
\begin{equation}
    R_{\rm out} \approx R_0 \bigg (\frac{M_\star}{M_\odot}\bigg)^\alpha,
    \label{eq::r_out}
\end{equation}
with $R_0 \approx 100\, \rm au$ and $\alpha = 0.5$ \citep{Andrews2013,Andrews2018}. To capture the observed flattening at very low masses, we adopt the modified form
\begin{equation}
     R_{\rm out} \approx R_0 \bigg (\frac{M_\star}{M_\odot}\bigg)^\alpha \bigg[ 1 + \bigg(\frac{M_{\rm break}}{M_\star}\bigg)^\beta \bigg]^{-\gamma},
     \label{eq::r_out_mod}
\end{equation}
where $M_{\rm break} \approx 0.1\, \rm M_\odot$ denotes the characteristic stellar mass at which the relation flattens \citep{Testi2016,Hendler2017}, and $\beta$, $\gamma$ parameterize the transition's sharpness and degree of flattening. In Figure~\ref{fig::disc_size} we overplot both Eqs.~(\ref{eq::r_out}) and (\ref{eq::r_out_mod}) on a compilation of observed brown dwarf disc radii; several systems fall below the simple power law but align with the modified relation. Applying Eq.~(\ref{eq::r_out_mod}) to 2M1510 AB ($M_\star = M_{\rm A} + M_{\rm B} = 0.0663\, \rm M_\odot$) yields $R_{\rm out} \approx 15\, \rm au$. Accounting for the fact that gas discs typically extend to several times the dust radius \citep{Ansdell2018,Sanchis:2021gd}, we infer a gas‐disc outer edge of $\sim 30\, \rm au$, marked by the dashed red line in Fig.~\ref{fig::ZKL_pcolor}. Such a compact primordial disc would be too small to undergo vZKL oscillations during polar alignment.

\begin{figure} \centering
\includegraphics[width=1\columnwidth]{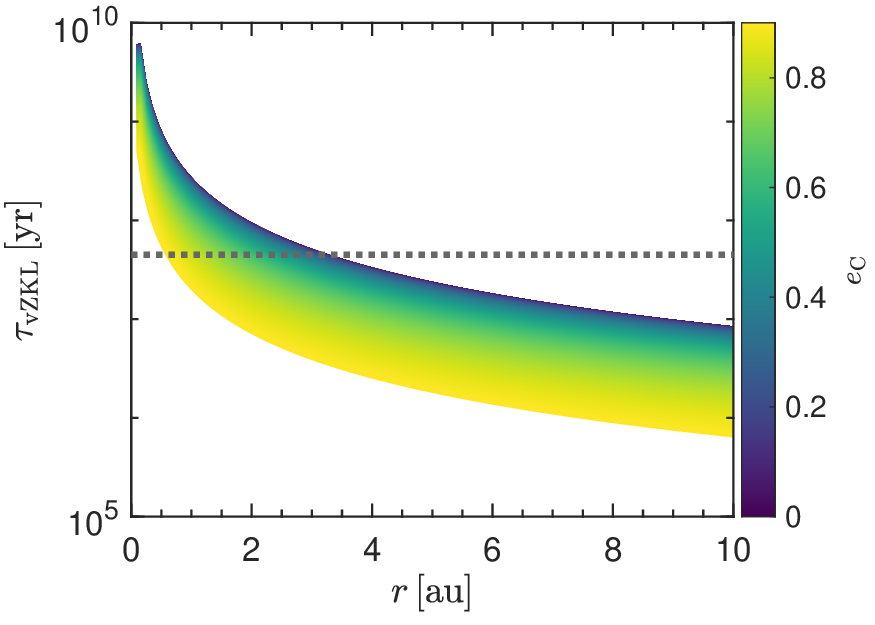}
\includegraphics[width=1\columnwidth]{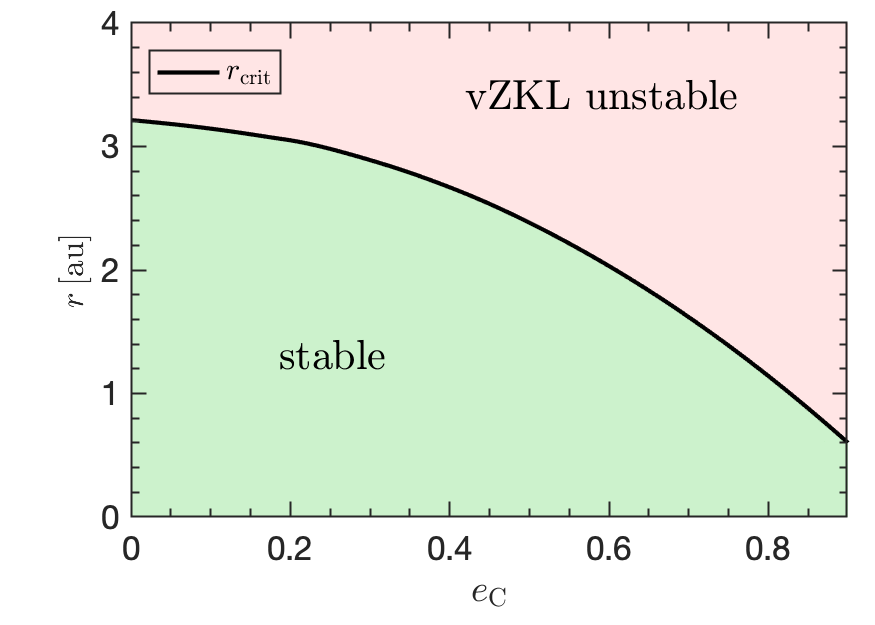}
\caption{ Stability of circumbinary particle orbits around 2M1510 AB under the influence of the tertiary component 2M1510 C. Top panel: The von Zeipel-Kozai-Lidov (vZKL) timescale, $\tau_{\rm vZKL}$ (in years), as a function of the particle's semi-major axis, $r$ (in au), from the central binary. The color bar indicates the eccentricity of the outer perturbing body, 2M1510 C, $e_{\rm C}$. The dotted horizontal line represents the estimated age of the 2M1510 system, $T_{\rm age} \approx 45\, \rm Myr$. Bottom panel: The critical radius, $r_{\rm crit}$, (in au) for vZKL instability as a function of eccentricity of 2M1510 C, $e_{\rm C}$. The critical radius is defined as the semi-major axis at which $\tau_{\rm vZKL} = T_{\rm age}$. The green shaded region below the black curve indicates orbits that are stable against vZKL oscillations ($r < r_{\rm crit}$), while the pink shaded region above the curve indicates orbits that are unstable to vZKL oscillations ($r > r_{\rm crit}$).}
\label{fig::KL}
\end{figure}

However, once the disc disperses, a planet may be susceptible to vZKL oscillations for certain values of the planet separation and eccentricity of 2M1510 C. We estimate the vZKL oscillation timescale for a circumbinary particle perturbed by the outer companion, 2M1510 C, defined as \citep{Kiseleva1998,Ford2000}:
\begin{equation}
    \tau_{\mathrm{vZKL}} = \frac{2\pi}{3}\bigg( \frac{P_3^2}{P} \bigg) \bigg( \frac{M_1+M_2+M_3}{M_3} \bigg) (1-e_3^2)^{\frac{3}{2}},
\end{equation}
where $P_3$ and $P$ are the orbital periods of 2M1510~C and a circumbinary test particle, respectively; $e_3$ is the eccentricity of 2M1510~C's orbit; and $M_1$, $M_2$, and $M_3$ are the masses of 2M1510~A, 2M1510~B, and 2M1510~C, respectively. For the mass of the outer companion, we assume $M_3 = M_1$. The top panel in Figure~\ref{fig::KL} illustrates that the vZKL timescale, $\tau_{\rm vZKL}$, is strongly dependent on the particle's separation, $r$, and the eccentricity of the perturber, $e_{\rm C}$. Specifically, $\tau_{\rm vZKL}$ decreases significantly with increasing r and increasing $e_{\rm C}$. This implies that particles at larger separations from the central binary or systems with a more eccentric outer perturber are susceptible to vZKL oscillations on shorter timescales. 

The bottom panel in Figure~\ref{fig::KL} delineates the boundary between stable and vZKL unstable orbits by plotting the critical radius, $r_{\rm crit}$, as a function of $e_{\rm C}$. This critical radius, where $\tau_{\rm vZKL}$ equals the system's age ($T_{\rm age} \approx 45\, \rm Myr$), decreases with increasing eccentricity of the outer companion (black curve). The green shaded area signifies orbits where particles are considered stable against vZKL oscillations. This stability implies that for particles with semi-major axes $r<r_{\rm crit}$, the timescale for significant vZKL-induced perturbations exceeds the age of the system. Conversely, the pink shaded region positioned above the black curve denotes orbits that are unstable due to vZKL oscillations. For particles in this zone, where $r < r_{\rm crit}$, the vZKL timescale is shorter than the system's age. As a result, these particles are susceptible to large-amplitude oscillations in their eccentricity and inclination. Therefore, much of the range of possible planetary-mass companions should correspond to stable polar circumbinary orbits.


\subsection{Planet formation in brown dwarf discs}
\label{sec::planet_formation}


\cite{Daemgen2016} found that approximately half of the brown dwarfs in their sample possess discs containing at least one Jupiter mass of material, confirming that many brown dwarfs have sufficient material to form Earth-mass to sub-Neptune planets. Grain growth and planetesimal formation via the streaming instability may be efficient in these discs \cite[e.g.,][]{Liu2020}.  \cite{Payne2007}  explored planet formation in brown dwarf discs by extending the sequential core accretion model of \cite{Ida2004,Ida2005} to lower-mass systems. Their findings indicate that discs with a total mass of a few Jupiter masses ($M_{\rm Jup}$) around a 0.05 $M_\odot$ brown dwarf cannot form giant planets, as rocky cores fail to accrete substantial gas envelopes. However, these discs can still produce rocky planets, with the maximum planet mass strongly dependent on the slope ($p$) of the dust’s radial surface density profile. Steeper profiles ($p=1.5$) enhance surface densities in the inner disc, enabling the formation of rocky planets with masses of 1--5 Earth masses ($M_{\oplus}$) at orbital distances of 0.2--3 au. In contrast, shallower profiles ($p=1.0$) reduce planet masses by a factor of 5--10 at similar distances \citep{Payne2007}.  Consequently, any giant planetary-mass companions observed around very low-mass objects like brown dwarfs are likely to have formed through alternative mechanisms, such as gravitational instability, as exemplified by 2M1207b \citep{Chauvin2005,Lodato2005}.

 As shown in Figures \ref{fig::alignment_time} and \ref{fig::align}, a disc within $\sim$100 $a_{\rm b}$ around 2M1510 AB is expected to align polar within its lifetime, potentially facilitating the formation of polar planets. Numerical simulations suggest that polar terrestrial planets can form efficiently around eccentric binaries \citep{Childs2021}, and studies of their long-term stability indicate that polar configurations are particularly stable around such systems \cite[e.g.,][]{Cuello2019,Chen2020}. Additionally, circumbinary discs can be more massive than circumstellar discs because the gravitational influence of two stars allows them to retain a larger reservoir of material \citep{Artymowicz1994,Bate1997}. A more massive disc can be more efficient in forming larger mass planets by providing higher surface densities and a greater amount of material, which enhances core accretion and gravitational instability processes, leading to the formation of more massive planets \citep{Lodato2004,Andrews2018}.

\subsection{Placing 2M1510 AB and VHS 1256 AB in a wider context}
The results shown in Figure \ref{fig::alignment_time} give a picture of alignment timescales across different binary parameters. Alignment to a polar configuration is more efficient for more massive binaries, for more eccentric binaries, and for smaller binary separations. Therefore, 2M1510\,AB would appear not ideal for efficient polar alignment, being low-mass and of moderate eccentricity and neither would VHS 1256\,AB due its wider separation. Binary brown dwarfs are rare, and planet-hosting brown dwarfs are rare too \citep[mainly due to detection difficulties;][]{blake_nirspec_2010,Gillon2013,Sahlmann2014}. While there is enough time for a disc of small radial extent to reach polar alignment around both VHS 1256\,AB and 2M1510\,AB, there remains a question about why polar circumbinary configurations have been identified in such exotic systems instead of more suitable and more common binaries. We speculate about three reasons: 
\begin{itemize}
    \item Detecting misaligned or polar planets in transit is difficult due to there not necessarily being a transit at every conjunction \citep{Schneider94,martin14, Chen2022}. All of the known transiting circumbinary planets were found in eclipsing binaries, which favours finding coplanar transiting planets \citep{Schneider94,DavidMartin2015}. A polar planet would only be expected to transit binaries if their line of apsides near the plane of the sky. With only 12 transiting circumbinary systems it is not surprising no polar transiting system has yet been found.
    \item There have been radial velocity surveys for circumbinary planets \citep{Konacki_2009,Martin_2019_BEBOP, Baycroft2024}. So far only one discovery of a circumbinary planet has been published from these \citep{Standing_2023, Baycroft2024}, and no dedicated search for retrograde apsidal precession has been performed to our knowledge. Most systems in the radial velocity samples have a low eccentricity \citep[\(e_b<0.3\);][]{Martin_2019_BEBOP}.
    \item Based on theoretical arguments, \cite{Cuello2019} predict the most likely polar circumbinary configuration ought to be found in equal-mass (\(q\approx1\)) binaries with moderate orbital eccentricities (\(e_b\approx0.4\)). They argue polar circumbinary planet formation capabilities are maximised for these parameters. 2M1510\,AB falls very much in this category, unlike the radial velocity binaries from the BEBOP survey \citep{Martin_2019_BEBOP,Baycroft2024} which are mostly \(q\lesssim 0.3\). VHS 1256\,AB does not fall so clearly in this category, the eccentricity being very high.
\end{itemize}

With 2M1510 (AB)b and VHS 1256 (AB)b, there are now two planetary mass companions on close-to-polar orbits around low-mass binaries. Delorme-1 is a visual, very low-mass binary, with an imaged planetary mass companion \citep{Delorme2013} which has an unconstrained mutual inclination at the moment \citep{Bowler2023} but could join these in the future. Combined with the lack of known polar planets orbiting more massive main-sequence binaries, this indicates that there is a potential mechanism to form polar objects specific to low-mass binaries. 

2M1510 (AB)b does not have mass and semi-major axis measurements yet. If it is comparable to VHS 1256 (AB)b, then they would both be long-period massive objects which would likely not have formed from a polar-aligning circumbinary disc (see Figure \ref{fig::alignment_time}). However 2M1510 (AB)b could well be a lower-mass close-in planet and hence have formed in a distinct way to VHS 1256 (AB)b. More observations to constrain the parameters of 2M1510 (AB)b are vital to test whether it falls into the same category as VHS 1256 (AB)b.

\section{Conclusions}
\label{sec::conclusion}

Recent observations by \cite{Baycroft2025} suggest that the retrograde precession of the binary brown dwarf system 2M1510\,AB is compatible with the presence of a polar circumbinary planet. If correct this would be the first detection of this type of orbital configuration. We investigated the evolution of a primordial circumbinary disc around the 2M1510\,AB brown dwarf binary to assess the likelihood that the disc could have evolved into a polar configuration.

From analytical calculations, we determined that the timescale for polar disc alignment decreases as the mass of the binary system increases. This suggested that discs surrounding more massive binary systems tend to align more rapidity compared to those around less massive systems. Furthermore, the efficient of alignment to a polar configuration in influences by several key factors: it is enhanced for more massive binaries, for binaries with higher orbital eccentricity, and for systems with smaller binary separations. When applied to brown dwarf binaries, which are characterized by their relatively low binary masses, the alignment timescales are expected to be significantly longer compared to more massive systems. However, within a specific parameter space of disc properties—such as disc size, viscosity, and initial misalignment—it is possible for the primordial disc around a brown dwarf binary, like 2M1510\,AB, to achieve polar alignment within the disc's lifetime. This suggests that while alignment processes are typically slower for low-mass binaries, certain conditions can still facilitate efficient polar alignment, even in systems like 2M1510\,AB.

We found that the critical tilt angle for a circumbinary disc to evolve into a polar alignment around 2M1510\,AB was determined to be $i_{\rm crit} \gtrsim 50^\circ$ for a low ratio of the disc angular momentum to the binary angular momentum. For a higher angular momentum ratio, the critical tilt converges to the Kozai-Lidov instability angle, $\sim 39^\circ$.  Thus, a primordial disc with an initial misalignment of $\gtrsim 50^\circ$ would be required for the disc to evolve into a polar configuration around 2M1510\,AB. We calculated a range of disc parameters, focusing on viscosity and aspect ratio, that facilitate the formation of a polar-aligned disc prior to its dispersal. For a disc viscosity of $\alpha = 10^{-4}$ and an aspect ratio of $H/r = 0.05$, a circumbinary disc with an outer edge extending to approximately $100a_{\rm b}$ can achieve polar alignment within the disc's lifetime. Using hydrodynamical simulations, we demonstrated that an initially inclined circumbinary disc evolves toward a polar configuration with respect to the orbit of 2M1510\,AB. A polar-aligned circumbinary disc could provide a favorable environment for the formation of polar circumbinary planets \citep[e.g.,][]{Smallwood2024b,Smallwood2024c}.


\section*{Acknowledgements}
We thank the anonymous referee for their constructive comments, which helped improve this manuscript. JLS acknowledges funding from the Dodge Family Prize Fellowship in Astrophysics. JLS is also supported by the Vice President of Research and Partnerships of the University of Oklahoma and the Data Institute for Societal Challenges. 
TAB and AHMJT acknowledge funding from the ERC/UKRI Frontier Research Guarantee programme (grant agreement EP/Z000327/1/CandY). RPN acknowledges funding from the STFC grant ST/X000931/1.

\section*{Data Availability}
The data supporting the plots within this article are available on reasonable request to the corresponding author. A public version of the {\sc phantom}, and {\sc splash} 
codes are available at \url{https://github.com/danieljprice/phantom}, \url{http://users.monash.edu.au/~dprice/splash/download.html}. 



\bibliographystyle{mnras}
\bibliography{ref.bib} 







\bsp	
\label{lastpage}
\end{document}